\documentclass[submission, copyright, creativecommons, svgnames]{eptcs}

\providecommand{\event}{PROLE 2016}

\usepackage{amsmath}

\usepackage{tikz}
\usetikzlibrary{arrows}
\usetikzlibrary{matrix}
\usepackage{listings}
\usepackage{amssymb}
\usepackage{xspace}
\usepackage{multirow}

\usepackage{wrapfig}

\usepackage{xcolor}

\definecolor{dkgreen}{rgb}{0,0.6,0}
\definecolor{gray}{rgb}{0.5,0.5,0.5}
\definecolor{mauve}{rgb}{0.58,0,0.82}
\definecolor{light-red}{rgb}{1.0,0.8,0.8}
\definecolor{yellow-orange}{rgb}{1.0,0.85,0.05}
\definecolor{gray-blue}{rgb}{0.57,0.72,0.84}
\definecolor{light-gray}{gray}{0.80}

\definecolor{DarkGreen}{rgb}{0.0, 0.2, 0.13}

\definecolor{apricot}{rgb}{0.98, 0.81, 0.69}
\definecolor{bananayellow}{rgb}{1.0, 0.88, 0.21}
\definecolor{babyblueeyes}{rgb}{0.63, 0.79, 0.95}

\usepackage{soul}
\sethlcolor{titlebgcolor}

\newcommand\red[1]{\setlength\fboxsep{0pt}\colorbox{light-red}{\strut #1}}

\newcommand\grey[1]{\setlength\fboxsep{0pt}\colorbox{babyblueeyes}{\strut #1}}

\newcommand\RL{reinforcement\ learning\xspace}
\newcommand\ML{machine\ learning\xspace}

\lstdefinestyle{cstyle}{
  escapechar={@},
  language=c,
  aboveskip=3mm,
  belowskip=3mm,
  showstringspaces=false,
  columns=flexible,
  basicstyle={\footnotesize\ttfamily},
  numbers=none,
  numberstyle=\tiny\color{gray},
  keywordstyle=\color{blue},
  keywordstyle=[2]\color{dkgreen},
  keywordstyle=[3]\color{magenta},
  commentstyle=\color{gray},
  stringstyle=\color{mauve},
  breaklines=true,
  breakatwhitespace=true,
  tabsize=3,
}

\lstdefinestyle{cstyleInline}{
  escapechar={@},
  language=c,
  aboveskip=3mm,
  belowskip=3mm,
  showstringspaces=false,
  columns=flexible,
  basicstyle={\footnotesize\ttfamily},
  numbers=none,
  numberstyle=\tiny\color{gray},
  keywordstyle=\color{blue},
  keywordstyle=[2]\color{dkgreen},
  keywordstyle=[3]\color{magenta},
  commentstyle=\color{gray},
  stringstyle=\color{mauve},
  breaklines=true,
  breakatwhitespace=true,
  tabsize=3,
  morekeywords={\#pragma}
}

\usepackage[
    textwidth=0.17\textwidth,
    textsize=scriptsize,
    colorinlistoftodos
    ,disable
    ]{todonotes}

\newcommand{\savespace}{\vspace*{-1em}}
\newcommand{\savespacebeforesection}{\vspace*{-1em}}
\newcommand{\savespaceaftersection}{\vspace*{-0.5em}}

\title{Towards Automatic Learning of Heuristics
      for Mechanical Transformations of Procedural Code\footnote{Work partially funded by EU FP7-ICT-2013.3.4 project 610686 POLCA, Comunidad de Madrid project S2013/ICE-2731 N-Greens Software, and MINECO Projects TIN2012-39391-C04-03 / TIN2012-39391-C04-04 (StrongSoft), TIN2013-44742-C4-1-R (CAVI-ROSE), and TIN2015-67522-C3-1-R (TRACES).}}
\author{Guillermo Vigueras \qquad Manuel Carro\footnote{Manuel Carro
    is also affiliated to the Universidad Politécnica de Madrid}
\institute{IMDEA Software Institute \\
             Campus de Montegancedo 28223 \\
             Pozuelo de Alarc\'on, Madrid, Spain}
\email{\{guillermo.vigueras,manuel.carro\}@imdea.org}
\and
Salvador Tamarit \qquad Julio Mari\~no
\institute{Universidad Polit\'ecnica de Madrid\\ 
	Campus de Montegancedo 28660\\
	Boadilla del Monte, Madrid, Spain}
\email{\{salvador.tamarit,julio.marino\}@upm.es}
}
\def\titlerunning{Towards Automatic Learning of Heuristics
      for Mechanical Transformations of Procedural Code}
\def\authorrunning{G. Vigueras, M. Carro, S. Tamarit \& J. Mari\~no, }

\begin{document}

%





\pagestyle{empty} 
\maketitle

\begin{abstract}
  The current trends in next-generation exascale systems go towards
  integrating a wide range of specialized (co-)processors into
  traditional supercomputers.  Due to the efficiency of heterogeneous
  systems in terms of Watts and FLOPS per surface unit, opening the
  access of heterogeneous platforms to a wider range of users is an
  important problem to be tackled.  However, heterogeneous platforms
  limit the portability of the applications and increase development
  complexity due to the programming skills required.
%
  Program transformation can help make programming heterogeneous
  systems easier by defining a step-wise transformation process that
  translates a given initial code into a semantically equivalent final
  code, but adapted to a specific platform.  Program transformation
  systems require the definition of efficient transformation
  strategies to tackle the combinatorial problem that emerges due to
  the large set of transformations applicable at each step of the
  process.
%
  In this paper we propose a \ML-based approach to learn heuristics to
  define program transformation strategies.  Our approach proposes a
  novel combination of \RL and classification methods to efficiently
  tackle the problems inherent to this type of systems. Preliminary
  results demonstrate the suitability of this approach.
\end{abstract}


\noindent\textbf{Keywords:}
Program Transformation, Machine Learning, Heterogeneous Systems

\savespacebeforesection
\section{Introduction}
\savespaceaftersection


The introduction of multi-core processors marked the end to an era
dominated by a constant increase of clock-rate, following Moore's law. Today, even the strategy to increase performance through higher core-counts is facing physical limitations. In addition, the performance increase results in power consumption issues which are currently one of the main limitations for the transition from petascale systems to next-generation exascale systems. Hence, the metric for measuring performance improvement of computational platforms has changed from absolute FLOPS numbers to FLOPS per Watt and FLOPS per surface unit of hardware. 

Thus, rather than having CPU-based supercomputers a growing trend goes
towards integrating a wide range of specialized (co-)processors into
traditional supercomputers. These specialized architectures can
outperform general purpose platforms while requiring lower energy
consumption and less real estate. A study of energy consumption in
data centers~\cite{Koomey2008} has revealed the impact that efficient
computer platforms can have in the worldwide energy consumption. This
study analyzed historical energy consumption data for the periods
2000-2005 and 2005-2010. The study results showed that in 2005 data
centers used 0.97\% of the total electricity consumption of the world,
while in 2010 this value increased to 1.3\% instead of the expected 2.2\%, according to the trends from the period 2000-2005. The study explained the reduction from 2.2\% down to 1.3\% as a consequence of the increasing use of energy-efficient computational platforms, showing the economic impact of using \textit{greener} architectures.


 However, integrating different specialized devices increases the degree of heterogeneity of the system. On the other hand, compilers typically
 optimize code for only one (homogeneous) destination infrastructure
 at a time, requiring the developer to manually adapt the program in order to be executed on hybrid architectures. As a result, the high degree of heterogeneity
 limits programmability of such systems to a few
 experts, and significantly reduces portability of the application onto
 different resource infrastructures.

\begin{figure} 
\centering
\begin{minipage}{0.7\linewidth}
\begin{lstlisting}[style=cstyle,caption=Initial code.,label=lst:convInitial]
// ABSTR: [3, 0, 0, 0, 0, 0, 1, 0, 0, 1, 1, 3, 2, 4, 0]

int dead_rows = K / 2;
int dead_cols = K / 2;

int normal_factor = K * K;

for (r = 0; r < N - K + 1; r++) {
  for (c = 0; c < N - K + 1; c++) {
    sum = 0;
    for (i = 0; i < K; i++) {
      for (j = 0; j < K; j++) {
        sum += input_image@\red{[r+i][c+j]}@ * kernel[i][j];
      }
    }
    output_image[r+dead_rows][c+dead_cols] = (sum / normal_factor);
  }
 }
\end{lstlisting}
%
%
\end{minipage}
\end{figure}

Opening the access of heterogeneous platforms to a wider spectrum of
users is an important issue to be tackled. In past decades, different
scientific areas showed an enormous advance and development thanks to
the use of massive parallel architectures to implement
\textit{in-silico} models. The use of ``virtual labs" through
computational platforms allowed a better understanding of the physical phenomena under study, investigate a much wider range of solutions, and drastically reduce the cost with respect to performing real experiments. The computational requirements of scientific and industrial applications are pushing a driving the development of exascale systems. For that reason, new programming models, tools and compilers are required in order to open the access of exascale systems to a wider programming community, enabling translational research to enhance human welfare.


In order to ease the programmability of heterogeneous platforms, we
have
proposed~\cite{tamarit15:padl-haskell_transformation,tamarit16:transformation-prole}
a rule-based transformation system which can help developers to
convert architecture-neutral code into code which can be deployed onto
different specific architectures. While the mentioned transformation
system is capable of performing sound program transformations, it
lacked an automatic method to guide the transformation process. This
paper describes a \ML-based approach to learn transformation
strategies from existing examples and to guide that system by using
these strategies.

Rule-based transformation systems pose different problems, like the
explosion in the number of states to be explored, arising from the
application of transformation rules in an arbitrary order or the
definition of a stop criteria for the transformation system. For the
latter we propose the use of classification trees and for the former
we propose a novel approach based on \RL. Both approaches use code
abstractions to effectively capture relevant features of a given
code. In order to illustrate the approach,
Listing~\ref{lst:convInitial} shows  code performing a 2D convolution
with its associated abstraction at the top of the listing (see
Section~\ref{sec:abstraction} for a description of the code
abstraction). The code in Listing~\ref{lst:convInitial} is well suited
to be parallelized, by adding OpenMP pragmas, for a multi-core
CPU. However, different code transformations, altering different code
abstractions features, are required to ease the translation to
different target platforms like GPUs or FPGAs.  For example, by  coalescing the two outer loops, obtaining a linear iteration space, or by transforming the data layout of 2D arrays into 1D arrays, we obtain a sequential code easier to map onto the two platforms mentioned before (see Section~\ref{sec:example}). The code features to be transformed are represented in the abstraction vector in Listing~\ref{lst:convInitial} through the first element (maximum loop depth), the twelfth element (number of non-1D arrays) and the fourteenth element (total number of loops). Our approach learns different sequences of code transformations that change the relevant abstraction features and obtain, for the example in Listing~\ref{lst:convInitial}, a code performing a 2D convolution within a \texttt{for} loop with a linear iteration space, using 1D arrays. We have performed a preliminary evaluation of the approach obtaining promising results that demonstrate the suitability of the approach for this type of transformation systems.


The rest of the paper is organized as follows: Section~\ref{sec:SoA} reviews previous work in the field of program transformation systems in general and previous approaches using \ML techniques. Section~\ref{sec:toolchain} describes the toolchain where our \ML-based approach is integrated. Section~\ref{sec:abstraction} describes the abstraction of code defined in order to apply \ML methods. Section~\ref{sec:Learning} describes the methods used to learn program transformation strategies. Section~\ref{sec:Results} presents some preliminary results and, finally, Section~\ref{sec:conclusions} summarizes the conclusions and proposes future work.

\savespacebeforesection
\section{State of the Art}
\label{sec:SoA}
\savespaceaftersection
Rule-based program transformation systems support the formulation of basic code transformation steps as generic rules and arrange their automatic application. This scheme offers the flexibility of splitting complex code transformations into small steps, admitting efficient implementations that can scale to large programs.  By adding more rules, the transformation system can also increase its capabilities in a scalable way. Rule-based systems also allow to decouple the definition of transformation steps (i.e. rule applications) from the strategies followed to apply the rules. This decoupling provides more flexibility to try different transformation strategies and select the best one according to the purpose of the system~\cite{Bagge03CodeBoost,Schupp2002}. Rule-based transformation has been used before to generate code for different computational platforms.

The transformation of C-like programs and its compilation into synchronous architectures, like FPGAs, and asynchronous platforms, like multi-core CPUs, has been addressed before~\cite{Brown2005-tr-opt_trans_hw}. However, the input language of this approach (Handel-C) is meant to specify synchronous systems, thus limiting its applicability to this type of systems.
A completely different approach is to use linear algebra to transform the
mathematical specification of concrete scientific
algorithms~\cite{Franchetti2006,Fabregat2013,DiNapoli2014}.
Here, the starting point is a mathematical formula and, once the formula is
transformed, code is generated for the resulting expression. 
However, a reasonable acceleration over hand-tuned code 
happens 
only for algorithms within that concrete domain, and applying the ideas to
other contexts does not seem straightforward.

Machine learning techniques have been already used in the field of compilation and program transformation~\cite{Mariani:2014,Pekhimenko:2010,Agakov:2006}. All these approaches share the same principles as they obtain an abstract representation of the input programs in order to apply \ML methods. Nevertheless, previous approaches target some specific architectures limiting the applicability of the approach and making it not suitable for heterogeneous platforms. In our approach, we obtain program abstractions in order to enable the \ML-guided transformation and compilation of a program written in C for heterogeneous systems. Additionally, none of the previous works have explored the use of \RL 
methods~\cite{RLSurvey1996} in the field of program transformation and compilation.

\savespacebeforesection
\section{Program Transformation Toolchain for Heterogeneous systems}
\label{sec:toolchain}
\savespaceaftersection
In this work we propose the automatic learning of transformation strategies for a rule-based transformation toolchain~\cite{tamarit15:padl-haskell_transformation,tamarit16:transformation-prole}. This type of systems offer a higher flexibility over compilation infrastructures where the code transformations and optimizations are embedded in the compilation engine~\cite{Stallman2009}. 

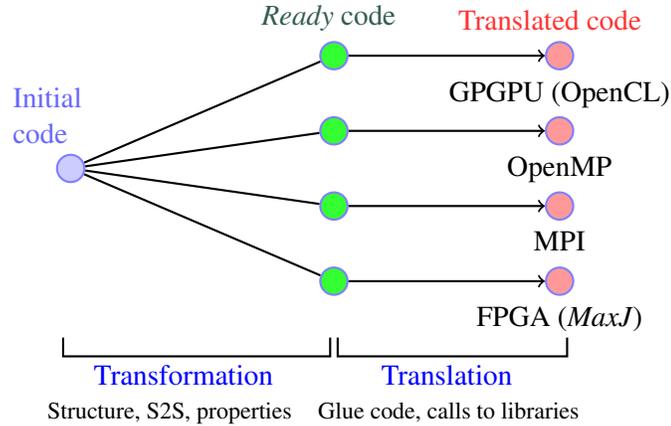
\begin{figure} 
  \begin{center}
\begin{minipage}{0.5\textwidth}
\begin{tikzpicture}[
    place1/.style={circle,draw=blue!50,fill=blue!20,thick},
    place2/.style={circle,draw=blue!50,fill=green!80,thick},
    place3/.style={circle,draw=blue!50,fill=red!40,thick},
    thick]
    \node (original) at (-1.5,0) [place1,label=above:\parbox{4em}{\textcolor{blue!60}{Initial code}}] {};
    \node (rgpu) at (2,1.5)
    [place2,label=above:\parbox{5em}{\textcolor{DarkGreen!80}{\emph{Ready} code}}] {};
    \node (romp) at (2,0.5) [place2] {};
    \node (rmpi) at (2,-0.5) [place2] {};
    \node (rfpga) at (2,-1.5) [place2] {};
    \node (gpu) at (5,1.5) [place3,label=below:GPGPU (OpenCL),label=above:\parbox{7em}{\textcolor{red!80}{Translated code}}] {};
    \node (omp) at (5,0.5) [place3,label=below:OpenMP] {};
    \node (mpi) at (5,-0.5) [place3,label=below:MPI] {};
    \node (fpga) at (5,-1.5) [place3,label=below:FPGA (\emph{MaxJ})] {};
    \draw [->] (original) -- (rgpu) -- (gpu);
    \draw [->] (original) -- (romp) -- (omp);
    \draw [->] (original) -- (rmpi) -- (mpi);
    \draw [->] (original) -- (rfpga) -- (fpga);
    \draw (-1.6,-2.25) -- (-1.6,-2.50) -- (1.95, -2.50) -- (1.95, -2.25);
    \draw (2.05,-2.25) -- (2.05,-2.50) -- (5.1, -2.50) -- (5.1, -2.25);
    \node [blue] at (0,-2.75) {Transformation};
    \node [blue] at (3.5,-2.75) {Translation};
    \node at (0.5,-3.25) {\parbox{12em}{\footnotesize\flushleft Structure, S2S, properties}};
    \node at (4.1,-3.1) {\parbox{12em}{\footnotesize\flushleft Glue code, calls to libraries}};
  \end{tikzpicture}
\end{minipage}
  \end{center}
  \caption{Stages of the transformation tool.}
  \label{fig:ana-trans-tool}
\end{figure}

The rule-based transformation toolchain mentioned before~\cite{tamarit15:padl-haskell_transformation,tamarit16:transformation-prole} defines a
 process consisting of two main stages
 (Fig.~\ref{fig:ana-trans-tool}). In the one, input C code is transformed into 
semantically equivalent C code which is better suited for a given
platform. It 
basically reshapes the input code taking into account syntactic/semantic
restrictions of the compiler/programming models of each target
platform (e.g.\ if the target compiler does not accept general
\texttt{\textbf{if-then-else}} statements, the input code will be transformed to remove them if possible, without modifying the program semantics). 
In this phase, transformation rules, written in a C-like DSL and working at the  \emph{abstract syntax tree} (AST) level, 
rewrite input code which matches a pattern and fulfills some semantic conditions,  into resulting code which is expressed as a template in the rewriting rule.
Since at each step several rules can generally be applied at several code locations, 
this first phase defines a search space exploration problem where the
states  correspond to the different configurations of code obtained from the application of rules, and the transitions among states correspond to the  applied rules.  The usually large number of applicable rules and code locations where they can be applied causes a combinatorial explosion of the number of states.
 
The second phase of the toolchain consists of a translation process
which generates code for a given platform taking as input the code
generated in the first phase.  This platform-specific code is fed to a
platform-specific compiler. This second phase poses another problem:
determine when code being subject to the transformation process is
ready for the translation phase, i.e.\ when the transformation process
can stop. Since some platforms might require a long compilation process (taking up to tens of hours for some FPGA compilers, for example, for moderate-sized programs), the toolchain should translate and compile only code that is likely to be successfully compiled on a given target platform, rather than trying to compile every single code obtained during the transformation phase.  Additionally, when considering the first and second phases together, the intermediate code states obtained in the transformation sequences of the first phase might not improve monotonically non-functional metrics (e.g. execution time, energy consumption, etc.), but still produce code with better performance in the second phase.

Given the nature of the problems associated with the transformation toolchain, we propose the use of different \ML techniques to effectively guide the transformation process.   Concretely, we have used \RL methods to learn transformation strategies that take into account the non-monotonic behavior of the transformation sequences, and we have used classification methods to learn appropriate states that can be translated and compiled in the target platforms. Machine learning techniques require descriptions of the problem domain; therefore, we compute an abstract representation of the input programs and the intermediate code generated during the transformation process.



%
%
%

\savespacebeforesection
\section{Mapping Code to Abstraction}
\label{sec:abstraction}
\savespaceaftersection

As mentioned in before, \ML operates on descriptions of the problem domain (C code, in our case).  We use code abstractions as descriptions; these should be rich enough to reflect the changes performed by the transformation rules at the AST level. For that reason, we include in the abstractions quantitative descriptions involving features like AST patterns, control flow, data layout, data dependencies, etc.  The current code abstraction consists of a vector of  features (described later) which capture traits which are modified by the transformation rules and also reflect code patterns that match the syntactic/semantic restrictions of target compiler/programming models.

These abstractions are generated through a code analysis tool that parses the AST to extract the features mentioned before,  implementing a function

\[A:\ Code \rightarrow Abstraction\]

\noindent
that maps codes to abstractions.  The tool has been implemented in Python using the \textit{pycparser}\footnote{\url{https://github.com/eliben/pycparser}} module and can extract feature information either by directly analyzing the code or by parsing code annotations provided by the user or by third-party analyzers.  These match those supported by the transformation engine~\cite{tamarit15:padl-haskell_transformation}:


\begin{itemize}
\item{\textbf{Maximum nested loop depth:}}  Depth of nested \texttt{for} loops. A value of $0$  means that no nested loops are present,  $1$ means that there are two nested loops,  and so on.

\item{\textbf{Number of function calls:}} Number of function calls present in the analyzed code.

\item{\textbf{Number of array writes shifted in bodies of \texttt{for} loops:}} This feature counts the number of array accesses with positive offset according to the following pattern: a \texttt{for} loop which performs multiple write accesses to the same array in the same iteration, some (but not all) of which use a positive offset with respect to the loop iteration variable.

Listing~\ref{lst:arr_wrt_shf} illustrates this feature.  It would take a value of 1 for the code on the left,  since for each iteration there are multiple write accesses to \texttt{v[]} and one of them has positive offset (because of the index \texttt{i+1}).   For the code on the right, this feature takes a value 0, since for each iteration only position \texttt{i+1} is written to: there is no position written to with a non-positive offset.  Detecting writes with this characteristic is necessary since this pattern would fail to  compile directly for a FPGA. 

\begin{figure}
  \begin{minipage}{0.55\textwidth}
\begin{lstlisting}[style=cstyleInline,caption=Array writes shifted.,label=lst:arr_wrt_shf,xleftmargin=.09\textwidth]
for(i=1;i<N;i+=2) {         for(i=0;i<N-1;i++) {
   v[i] = v[i-1];              aux = i*i;
   v[i+1] = v[i-1]*i;          v[i+1] = aux;
}                           }
\end{lstlisting}
\end{minipage}
\hfill
\begin{minipage}{0.45\textwidth}
\begin{lstlisting}[style=cstyleInline,caption=Static loop limits.,label=lst:loop_stc_lim,xleftmargin=.13\textwidth]
for(j=0;j<N;j++) {     
   for(i=0;i<size(v);i++)
      update(v,i);
   clean(v);
}
\end{lstlisting}
\end{minipage}
\end{figure}

\item{\textbf{Irregular loops:}} This feature takes a binary value and
  informs whether there are loops containing or not statements that
  alter the regular flow of a structured program (e.g., there are
  \texttt{break} or \texttt{continue} statements).

\item{\textbf{Global variables:}} This binary feature states if any
  global variable is written or not within the fragment of code of interest.

\item{\textbf{\texttt{if} statements:}} This feature counts the number of \texttt{if} statements within the piece of code analyzed.

\item{\textbf{Static limits of \texttt{for} loops:}} This feature captures whether some \texttt{for} loop in the analyzed code has a non-static iteration limit, i.e. if it can change between executions of the loop.
Listing~\ref{lst:loop_stc_lim} shows an example where the upper limit of the innermost loop is non-static, since it might change for each iteration of the outer loop if elements of \texttt{v} are removed by function \texttt{clean}. 

%

\item{\textbf{Iteration independent loop:}} This feature counts the number of \texttt{for} loops in the analyzed code which do not have carried dependencies across iterations.  This information is obtained from annotations provided by the user or by external code analyzers.

\item{\textbf{Any \texttt{for} loop with \texttt{loop\_schedule}:}} This feature detects patterns where two nested loops are used to iterate over an array split in fragments.  This code pattern is detected through an annotation that can be inserted either by the user or by the transformation tool after applying a loop scheduling transformation. Listing~\ref{lst:scheduled_loop} shows an example of \texttt{loop\_schedule} where array \texttt{v} is accessed in chunks of size \texttt{N}. 

  \begin{figure}
\begin{minipage}[t]{0.45\textwidth}
\begin{lstlisting}[style=cstyleInline,caption=Loop with schedule pattern.,label=lst:scheduled_loop,xleftmargin=.13\textwidth]
#pragma stml loop_schedule
for(j=0; j<M; j++) {     
   w[j] = 0;
   for(i=0;i<N;i++)
      w[j] += v[j*N+i];
}
\end{lstlisting}
\end{minipage}
\hfill
\begin{minipage}[t]{0.45\textwidth}
\begin{lstlisting}[style=cstyleInline,caption=Aux. variable array index.,label=lst:arr_aux_var,xleftmargin=.13\textwidth]
aux = 0;
for(j=0; j<N; j++) {     
   w[j] = v[aux];
   aux++;
}
\end{lstlisting}
\end{minipage}
  \end{figure}


\item{\textbf{Number of loop-invariant variables:}} This feature quantifies the number of variables that are assigned outside a \texttt{for} loop and are not modified inside it.

\item{\textbf{Number of loop hoisted variable modifications:}} This feature is used to count the number of variables that are assigned values outside a \texttt{for} loop and are modified within the loop.

\item{\textbf{Number of non-1D array:}} This feature counts the number of arrays in the code with two or more dimensions.

\item{\textbf{Number of auxiliary variables to access arrays:}} This feature counts the number of auxiliary variables used to index an array. If a variable used as array index is not one of the iteration variables of the \texttt{for} loops that iterate over the array, then it is considered as an auxiliary index variable. Listing~\ref{lst:arr_aux_var} shows a simple example where the \texttt{aux} variable is used to index the array \texttt{v} instead of using the iteration variable \texttt{j} of the \texttt{for} loop.


\item{\textbf{Total number of \texttt{for} loops:}} This feature counts the total number of \texttt{for} loops, either nested or non-nested. For example, this feature takes the value 2 for the code in Listing~\ref{lst:scheduled_loop}.

\item{\textbf{Non-normalized \texttt{for} loops:}} This feature counts the number of non-normalized \texttt{for} loops.  A \texttt{for} loop is considered  normalized iff its iteration step is $1$.

%
%
%
\end{itemize}

The features described above were sufficient to obtain preliminary results for a set of use cases (Section~\ref{sec:Results}). However, we plan to increase the vector of features as the set of rules and use cases grows.

\savespacebeforesection
\section{Automatic Learning of Transformation Heuristics}
\label{sec:Learning}
\savespaceaftersection
%

As mentioned before, two outstanding problems faced by the
transformation engine are dealing with efficiently
finding transformation sequences in a very large (even infinite) state space with a
non-monotonic fitness function and deciding when to stop searching,
since there is no a-priori general way to determine that an optimum
(in the sense of the most efficient possible code) state has been
reached.
%
Our approach uses classification trees to solve the latter and \RL to solve the former.  We  will describe our approaches in the next sections.

\subsection{Classification Trees}
\label{sec:CT}

In \ML and statistics, classification is the problem of identifying the category to which a new observation belongs among a set of pre-defined categories.  Classification is done on the basis of a training set of data containing observations for which it is known to which category they belong~\cite{Marsland2009}. Different approaches can be used to classify.  We have decided to start evaluating the adequacy of classification trees for our problem since it performs feature selection without complex data preparation.

A classification tree is a simple representation to classify examples
according to a set of input features. The input features should have
finite discrete domains and there is a single target variable called
the \textit{classification} feature.  Each element of the domain of
the target variable is called a class.  In a classification tree each
internal (non-leaf) node is labeled with an input feature and each
leaf node is labeled with a class or a probability distribution over
the classes. Thus, a tree can be learned by splitting the source data
set into subsets based on values of input features. This process is
recursively repeated on each derived subset. The recursion is completed when the subset of data in a node has the same value for the target variable, or when splitting no longer improves the predictions. The source data typically comes in records of the form 

\[(\textbf{x},Y) = ([x_1, x_2, x_3, ..., x_k], Y)\]

The dependent variable, $Y$, is the target variable that the classification tree generalizes in order to be able to classify new observations. The vector $\textbf{x}$ is composed of the input features $x_i$, used for the classification. In this context, the source input data for our problem is composed of the vectors of the abstractions described in Section~\ref{sec:abstraction}, i.e., $k=15$ in our case. The domain of the target variable can take values among the four different final platforms we currently support: FPGA, GPU, Shared-Memory CPU (SM-CPU) and Distributed-Memory CPU (DM-CPU). Since a given code and its associated abstraction might be well suited for more than one platform, we compute the number of classes for our target variable as $2^{n}-1$, where $n$ is the number of target platforms. Thus, we obtain $2^{4}-1=15$ classes.

%

The classes obtained for the target variable make it possible to
define the final states of the transformation stage of the toolchain
described in Section~\ref{sec:toolchain}, which will also serve to
define the final states for the \RL algorithm that is described
next. The classification-based learning described in this section has
been implemented using the Python library \textit{Scikit-learn}~\cite{scikit2011}. This library implements several \ML algorithms, provides good support and ample documentation, and is widely used in the scientific community.

\subsection{Reinforcement Learning}
\label{sec:RL}

Reinforcement learning~\cite{Marsland2009} is an area of \ML
 concerned with how software agents ought to take actions in an environment to maximize some notion of cumulative reward.  A \RL agent interacts with its environment in discrete time steps. At each time $t$, the agent receives an observation $o_t$ which typically includes the reward $r_t$. It then chooses an action $a_t$ from the set of available actions, which is subsequently sent to the environment. The environment moves from the current state $s_t$ to a new state $s_{t+1}$ providing the reward $r_{t+1}$ associated with the transition $(s_t,a_t,s_{t+1})$. The goal of a \RL agent is to collect as much reward as possible. 

According to the previous description, \RL seems well suited to represent the optimization process of a programmer or a compiler, since it typically consists of
iteratively improving  an initial program in discrete steps, where code changes correspond to actions and code versions obtained during the optimization process correspond to states.  Moreover, code is typically evaluated after every change, often  according to some non-functional properties such as execution time, memory consumption speedup factor, \ldots  The result of these evaluations
can be easily translated into rewards and penalties that support the
learning process.

\begin{figure*}[!b]
   \centering

\hspace*{4em}
\begin{tikzpicture}[mycell/.style={minimum size=1cm}]

\draw (-2,-3) -- (3,-3);
\draw (-2,-2) -- (3,-2);
\draw (-2,0) -- (3,0);
\draw (-2,1) -- (3,1);
\draw (-2,2) -- (3,2);
\draw (-2,2) -- (-2,-3);
\draw (-1,2) -- (-1,-3);
\draw (0,2) -- (0,-3);
\draw (2,2) -- (2,-3);
\draw (3,2) -- (3,-3);
\matrix (A) [matrix of math nodes,row sep=-\pgflinewidth, column
sep=-\pgflinewidth,nodes={mycell}] {
    & r0      & r1 &        &        & rn \\
\node (s0) {s0}; & q_{0,0}      & \textcolor{blue}{\ensuremath{\mathbf{q_{0,1}}}} & \cdots & \cdots & q_{0,n} \\
\node (s1) {s1}; & q_{1,0}      &  q_{1,1} &      &        & q_{1,n} \\
    & \vdots &   &        &        & \vdots \\
    & \vdots &   &        &        & \vdots \\
\node (sm) {sm}; & q_{0,m}      & q_{1,m} & \cdots & \cdots & q_{m,n} \\
};
\node (c0) [left of=s0,xshift=-15] {$\mathcal{C}_0$};
\node (c1) [left of=s1,xshift=-15] {$\mathcal{C}_1$};
\node (cm) [left of=sm,xshift=-15] {$\mathcal{C}_m$};
\draw [->] (c0) -- ++(1.25,0) node [midway,above] {\scriptsize SM};
\draw [->] (c1) -- ++(1.25,0) node [midway,above] {\scriptsize SM};
\draw [->] (cm) -- ++(1.25,0) node [midway,above] {\scriptsize SM};

\node (rc0) [left of=c0, yshift=20,xshift=-25] {$c_0$};
\node (rc1) [below of=rc0,yshift=17] {$c_1$};
\node (rc2) [below of=rc1,yshift=17] {$c_2$};
\draw [->] (rc0) -- (c0);
\draw [->] (rc1) -- (c0);
\draw [->] (rc2) -- (c0);
\node [right of=rc0,xshift=-7,yshift=5] {\scriptsize A};

\node (rc3) [left of=c1, yshift=10,xshift=-25] {$c_3$};
\node (rc4) [below of=rc3,yshift=17] {$c_4$};
\node (rc5) [below of=rc4,yshift=17] {$c_5$};
\draw [->] (rc3) -- (c1);
\draw [->] (rc4) -- (c1);
\draw [->] (rc5) -- (c1);

\node (rc6) [left of=cm, yshift=10,xshift=-25] {$c_6$};
\node (rc7) [below of=rc6,yshift=17] {$c_7$};
\node (rc8) [below of=rc7,yshift=17] {$c_8$};
\draw [->] (rc6) -- (cm);
\draw [->] (rc7) -- (cm);
\draw [->] (rc8) -- (cm);

\node [below of=rc5] {\vdots};
\node [below of=c1] {\vdots};

\draw (rc0.north west) rectangle (rc8.south east);
\node [above of=rc0] (ac) 
    {\parbox{4em}{\centering\footnotesize Concrete \\ code}};

\node [above of=c0, yshift=20]
    {\parbox{4em}{\centering\footnotesize Code \\ abstraction}};
\draw (c0.north west) rectangle (cm.south east);


\node (pr) [color=gray,right of=A-4-6,xshift=10]
    {\resizebox{5mm}{4cm}{\}}};
\node [xshift=165,yshift=-12] (right of=pr) {\parbox{9em}{States $\approx$ \\ Code Abstractions}};

\node (pu) [color=gray,above of=A-2-3,xshift=30,yshift=15]
    {\rotatebox{90}{\resizebox{5mm}{4cm}{\}}}};
\node  [xshift=15,yshift=100] (above of=pu) {Actions = Rules};

\end{tikzpicture}

   \caption{State-Action table for code, code abstraction, and rules.}
   \label{fig:sa-code-rules}
\end{figure*}
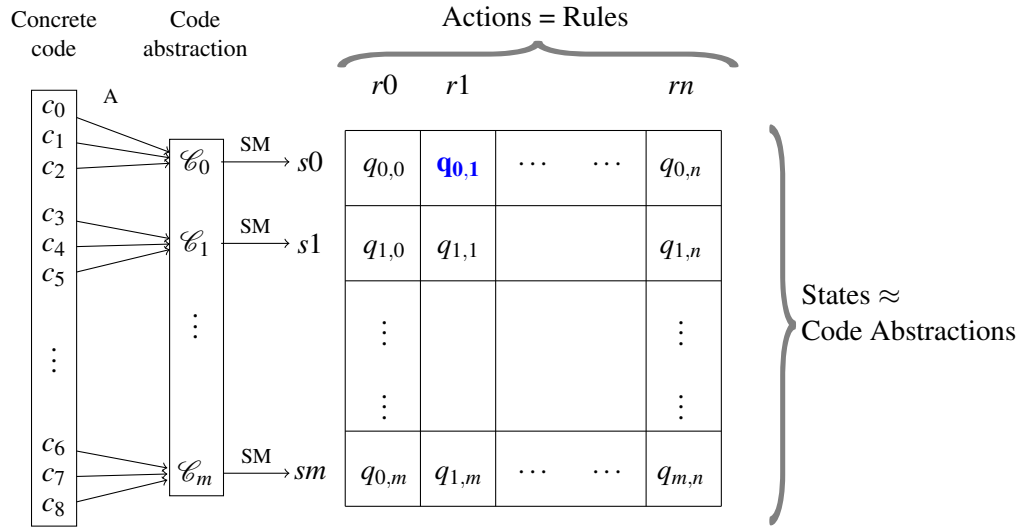

The result of the learning process of the agent is a \textit{state-action} table (Fig.~\ref{fig:sa-code-rules})
 which will eventually contain values for each combination $(s, a)$ of states and actions.   These values are scores for the expected profit to be obtained from applying action $a$ to state $s$.  This table is initially filled in with a default value and is iteratively updated following a learning process which we briefly describe below.


The process of \RL is based on a set of predetermined transformation sequences which are assumed to be models to learn from.  Each sequence $S$ is composed of a set of states $S = s_0, s_1, \ldots, s_{t-1}, s_t, s_{t+1}, \ldots, s_l$ and the actions which transform one state into the next one.
The final state of each transformation sequence has a different reward
value which in our case is related with the performance of the final code
corresponding to state $s_l$ in the sequence (better performance gives
higher rewards).  The training phase of  \RL consists of an iterative, stochastic process in which states (i.e. code abstractions) from the training set are randomly selected. For each  state $s$ a \emph{learning episode} is started by selecting the action $a$ with the highest value in $Q$ for that $s$ and moving to a new state $s'$ according to transition $(s,a,s')$. From state $s'$ the same transition process is repeated to advance in the learning episode until a final state is reached or a given number of episode steps is performed. When the episode terminates, the values in $Q$  corresponding to the states and actions of the visited sequence are updated according to (\ref{eq:RL}), where $Q_{init}(s_t,a_t)$ is the initial value of $Q$ for state $s_t$ and action $a_t$.   Note that $s_t$ (resp.\ $a_t$) is not the $t$-th state in some fixed ordering of states, but the $t$-th state in the temporal ordering of states in the sequence used to learn.

\savespace
\begin{equation}
Q(s_t,a_t) = \left\{
\begin{array}{ll}
Q(s_t,a_t) + \alpha~\cdot~(r_{t+1}+ ~\gamma~\cdot~Q(s_{t+1}, a_{t+1})~-~Q(s_t,a_t)~) & $if $ s_t\ not\ final \\
                              & \\
             Q_{init}(s_t,a_t) & $otherwise$
\end{array}
\right.
\label{eq:RL}                                                                     \end{equation} 

\medskip

\begin{figure}
\centering
\begin{tabular}{ |c|c| } 
\hline
 $SM:\ Abstraction \rightarrow State$ \                      & $Q:\ State\ \times\ Action \rightarrow \mathbb{R}$     \\ \cline{2-2}
                         $AM:\ Rule \rightarrow Action$         &  $RS:\ Code\ \rightarrow Rule$                                    \\  \hline
\multicolumn{2}{ |c| }{$RS(c)\ =\ \operatorname*{arg\,max}_{ru \in Rule} Q(SM(A(c)),AM(ru))$} \\  
 \hline
\end{tabular}
\caption{RL function definitions.}
\label{tbl:RL_functions}
\end{figure}

The final states in (\ref{eq:RL}) are defined based on the classification described in Section~\ref{sec:CT}. Two additional parameters appear in (\ref{eq:RL}): the learning rate $\alpha, 0 < \alpha \le 1$, and the discount factor $\gamma, 0 < \gamma \le 1$. The learning rate determines to what extent the newly acquired information will override the old information. A factor of 0 will make the agent not learn anything, while a factor of 1 would make the agent consider only the most recent information.  The discount factor implements the concept of \textit{delayed reward} by determining the importance of future rewards. A factor of 0 will make the agent opportunistic by considering only current rewards, while a factor close to 1 will make it strive for long-term rewards.  If the discount factor reaches or exceeds 1, the values in $Q$  may diverge~\cite{Marsland2009}.

In order to use the \RL \textit{state-action} in our setting, 
we need to define some mappings.  The abstraction of a concrete piece
of code is provided by function $A$ (described in
Section~\ref{sec:abstraction}).  Code abstractions and transformation rules
must be mapped to states and actions, respectively, in order to index
the \textit{state-action} table.  This mapping is done through functions $SM$ and $AM$ (Fig.~\ref{tbl:RL_functions}). Based on the mapping of abstractions and rules defined, the \RL \textit{state-action} table of (\ref{eq:RL}) can also be modeled as a function $Q$ (see Fig.~\ref{tbl:RL_functions}).

Using functions $A$, $SM$, $AM$ and $Q$, the strategy of the transformation toolchain for selecting rules at each transformation step can be modeled with function $RS$, defined in Fig.~\ref{tbl:RL_functions}. This function takes as input a given code $c$ and selects the transformation rule $r$ associated to action $AM(r)$ that maximizes the value provided by $Q$ for the state $SM(A(c))$ associated to input code $c$. Thus, the rule selection strategy is modeled by function $RS$ defined in Fig.~\ref{tbl:RL_functions}.

The operator $\operatorname*{arg\,max}$ in function $RS$ may return, by definition, the empty set, a singleton, or a set containing multiple elements. However, in our problem, parameters $\alpha$ and $\gamma$ as well as the reward values $r_{t+1}$
 appearing in (\ref{eq:RL}) can be tuned to ensure that a single rule is returned, thus avoiding a non-deterministic $RS$ function. Section~\ref{sec:Results} gives further details on how we selected their values. 

The relationships among the code, its abstraction, the rules, and the
contents of the state-action matrix are depicted in
Fig.~\ref{fig:sa-code-rules}.  Table $Q$ is used as follows: for a
concrete code $c_k$ we find its abstraction $\mathcal{C}_i = A(c_k)$.
Let us assume $i=0$.  From the row $i$ corresponding to
$\mathcal{C}_i$ in matrix $Q$ we obtain the column $j$ with the
highest value $q_{i,j}$ (in our example, $q_{0,1}$, in blue and
boldface).  Column $j$ corresponds to rule $R_j$, which is expected to
give the most promising code when applied to a code state whose
abstraction is $\mathcal{C}_i$ (in our case it would be $R_1$).  Rule
$R_j$ would be applied to $c_k$ to give $c'$.  If $c'$ corresponds to
a final state, the procedure finishes.  Otherwise, we repeat the
procedure taking $c'$ as input and finding again a rule to transform
$c'$.

 In order to implement the \RL-based module in charge of learning
 heuristics for program transformation we have used the Python library
 \textit{PyBrain}~\cite{pybrain2010}. This library adopts a modular
 structure separating in classes the different concepts present in
 \RL, such as the environment, the observations and rewards, the actions, etc. This modularity allowed us to extend the different classes and ease their adaptation to our problem. The \textit{PyBrain} library also provides flexibility to configure the different parameters of the \RL algorithm.
 
\subsection{A Simple Example}
\label{sec:example}

We will use a 2D convolution kernel as an example to show the
resulting \textit{state-action} table obtained after learning from a
simple transformation sequence with five states and two transformation
rules. The first rule ($R_0$) considered in the example transforms a
non-1D array into a 1D array. The second rule ($R_1$) performs a
collapse of two nested \texttt{for} loops producing a single loop. The
initial, intermediate and final codes obtained from the application of
these rules is described below.  Listing~\ref{lst:convStep1} through
Listing~\ref{lst:convStep4} highlights changes in code using \red{this
  style} to indicate the portion of the code that will be changed
after rule application and  code highlighted using
{\color{mauve}\underline{this style}} indicates the code resulting
after applying a transformation rule.

\begin{figure}
\begin{minipage}[t]{0.48\textwidth}
\begin{lstlisting}[style=cstyle,caption=Transformation step 1.,label=lst:convStep1]
// ABSTR: [3, 0, 0, 0, 0, 0, 1, 0, 0, 1, 1, @{\color{mauve}\underline{2}}@, 2, 4, 0]

int dead_rows = K / 2;
int dead_cols = K / 2;

int normal_factor = K * K;

for (r = 0; r < N - K + 1; r++) {
  for (c = 0; c < N - K + 1; c++) {
    sum = 0;
    for (i = 0; i < K; i++) {
      for (j = 0; j < K; j++) {
        sum += input_image@{\color{mauve}\underline{[(r+i)*(N - K + 1) + (c+j)]}}@ * @\red{kernel[i][j]}@;
      }
    }
    output_image[r+dead_rows][c+dead_cols] = (sum / normal_factor);
  }
 }
\end{lstlisting}
\end{minipage}%
\hfill
\begin{minipage}[t]{0.48\textwidth}
\begin{lstlisting}[style=cstyle,caption=Transformation step 2.,label=lst:convStep2]
// ABSTR: [3, 0, 0, 0, 0, 0, 1, 0, 0, 1, 1, @{\color{mauve}\underline{1}}@, 2, 4, 0]

int dead_rows = K / 2;
int dead_cols = K / 2;

int normal_factor = K * K;

for (r = 0; r < N - K + 1; r++) {
  for (c = 0; c < N - K + 1; c++) {
    sum = 0;
    for (i = 0; i < K; i++) {
      for (j = 0; j < K; j++) {
        sum += input_image[(r+i)*(N - K + 1) + (c+j)] * @{\color{mauve} \underline{kernel[i*K+j]}}@;
      }
    }
    output_image@\red{[r+dead\_rows][c+dead\_cols]}@ = (sum / normal_factor);
  }
 }
\end{lstlisting}  
\end{minipage}

\resizebox{0.48\linewidth}{!}{
\begin{minipage}[t]{0.5\linewidth}
\begin{lstlisting}[style=cstyle,caption=Transformation step 3.,label=lst:convStep3]
// ABSTR: [3, 0, 0, 0, 0, 0, 1, 0, 0, 1, 1, @{\color{mauve}\underline{0}}@, 2, 4, 0]

int dead_rows = K / 2;
int dead_cols = K / 2;

int normal_factor = K * K;

@\red{for(r = 0; r < N - K + 1; r++)}@ {
  @\red{for(c = 0; c < N - K + 1; c++)}@ {
    sum = 0;
    for (i = 0; i < K; i++) {
      for (j = 0; j < K; j++) {
        sum += input_image[(@\red{r}@+i)*(N - K + 1) + (@\red{c}@+j)] * kernel[i*K+j];
      }
    }
    output_image@{\color{mauve}\underline{[(\red{r}+dead\_rows)*(N-K+1) +} \underline{(\red{c}+dead\_cols)]}}@ = (sum / normal_factor);
  }
 }
\end{lstlisting}
\end{minipage}}
\hfill
\resizebox{0.48\linewidth}{!}{%
\begin{minipage}[t]{0.5\linewidth}
\begin{lstlisting}[style=cstyle,caption=Transformation step 4.,label=lst:convStep4]
// ABSTR: [@{\color{mauve}\underline{2}}@, 0, 0, 0, 0, 0, 1, @{\color{mauve}\underline{1}}@, 0, 1, 1, 0, 2, @{\color{mauve}\underline{3}}@, 0]

int dead_rows = K / 2;
int dead_cols = K / 2;

int normal_factor = K * K;

@{\color{mauve} \underline{for (z = 0; z < (N - K + 1)*(N - K + 1); z++)}}@ {
  @{\color{mauve}\underline{int r = (z / (N - K + 1));}}@
  @{\color{mauve}\underline{int c = (z \% (N - K + 1));}}@
    sum = 0;
    for (i = 0; i < K; i++) {
      for (j = 0; j < K; j++) {
        sum += input_image[(r+i)*(N - K + 1) + (c+j)] * kernel[i*K+j];
      }
    }
    output_image[(r+dead_rows)*(N - K + 1) +
	       (c+dead_cols)] = (sum / normal_factor);
}
\end{lstlisting}
\end{minipage}}
\end{figure}

Listing~\ref{lst:convInitial} shows the initial code and the associated vector of features, as a code comment, 
according to the description in Section~\ref{sec:abstraction}. Listing~\ref{lst:convStep1} shows the result of applying rule $R_0$ to the code in Listing~\ref{lst:convInitial}.  The array \texttt{input\_image} is transformed into a 1-D array and the vector of features associated to code is changed accordingly. Listing~\ref{lst:convStep2} shows the result of applying again rule $R_0$ to the code in Listing~\ref{lst:convStep1}.  The array \texttt{kernel} is then transformed into a 1-D array and the feature vector is updated. Listing~\ref{lst:convStep3} shows the result of applying rule $R_0$ to the code in Listing~\ref{lst:convStep2}.  Again, the array \texttt{output\_image} is transformed into a 1-D array and the vector of features updated.
Listing~\ref{lst:convStep4} shows the result of applying rule $R_1$ to the code in Listing~\ref{lst:convStep3}.  The two outermost loops are collapsed into one \texttt{for} loop, but keeping an iteration space with the same number of iterations.  Note that the code abstraction reflects the change, since the number of loops has decreased by one.


\begin{figure} 
  \begin{center}
\begin{tabular}{ c|c|c||c| } 
\cline{2-4}
                                                            & $AM(R_0)$                    & $AM(R_1)$ & $RS(C_i)$ \\ \hline
 \multicolumn{1}{ |c| }{$SM(A(C_0))$} &  \grey{17.03718317} &  16.21544456 & $R_0$ \\
 \multicolumn{1}{ |c| }{$SM(A(C_1))$} &  \grey{17.25327145} & 16.80486418 & $R_0$ \\
 \multicolumn{1}{ |c| }{$SM(A(C_2))$} &  \grey{17.51541052} & 16.7189079 & $R_0$ \\
 \multicolumn{1}{ |c| }{$SM(A(C_3))$} &  16.72942327 & \grey{17.78007298} & $R_1$ \\
 \multicolumn{1}{ |c| }{$SM(A(C_4))$} &  1.                  &    1.                          & - \\  

 \hline
\end{tabular}
  \end{center}
\caption{Values learned for $Q$ table.}
\label{tbl:learningTable}
\end{figure}

Fig.~\ref{tbl:learningTable} shows a table with the resulting values
of the \textit{state-action} table ($Q$) for the transformation
sequence described before. Fig.~\ref{tbl:learningTable}
has as many rows as states obtained from the evaluation of
$SM(A(C_i))$ for each code $C_i$, where $C_0$ is the initial code and
$C_4$ is the final code classified as ready-code for an
FPGA. Fig.~\ref{tbl:learningTable} shows the learned sequence composed
of four steps: three consecutive applications of rule $R_0$ and one
application of rule $R_1$. The values in $Q$  for this sequence are
highlighted in \grey{blue}, and they decrease from
the state $SM(A(C_3))$, with the highest reward, down to the initial
state $SM(A(C_0))$. This decay behavior
is caused by the discount factor ($\gamma$) introduced in
(\ref{eq:RL}). It should be noted that the values in $Q$ for the final states are not updated by the recursive expression in (\ref{eq:RL}). Thus, the final state $SM(A(C_4))$ keeps the initial value of 1. It should be noted that numerical values in table shown in Fig.~\ref{tbl:learningTable} are obtained as the result of an iterative and stochastic learning process as defined in (\ref{eq:RL}) and using parameter values described in Section~\ref{sec:Results}.

The example used in this section shows the transformation of a piece of C code. However, the fact that the \ML methods used work on program abstractions makes the approach generic and suitable for other imperative languages (e.g., FORTRAN, which is also widely used in scientific computing). The application of the approach to other languages would require changes to the tool described in Section~\ref{sec:abstraction} in order to account for some specific syntactical patterns of a particular programming language. Nevertheless, most of the abstraction features identified and described in Section~\ref{sec:abstraction} are also applicable to other imperative languages since they capture common aspects like control flow, data layout, data dependencies, etc.

\savespacebeforesection
\section{Results}
\label{sec:Results}
\savespaceaftersection

We will present now some results obtained using
\ML-based transformation strategies.  We will first show evidence to
support our claim, made in Section~\ref{sec:RL}, regarding the
non-monotonic behavior of non-functional properties for transformation
sequences and, second, we will evaluate the applicability of reinforcement learning to learn from these non-monotonic sequences.

\begin{figure}
\centering
\begin{minipage}{0.45\textwidth}
\centering
\includegraphics[width=0.9\textwidth]{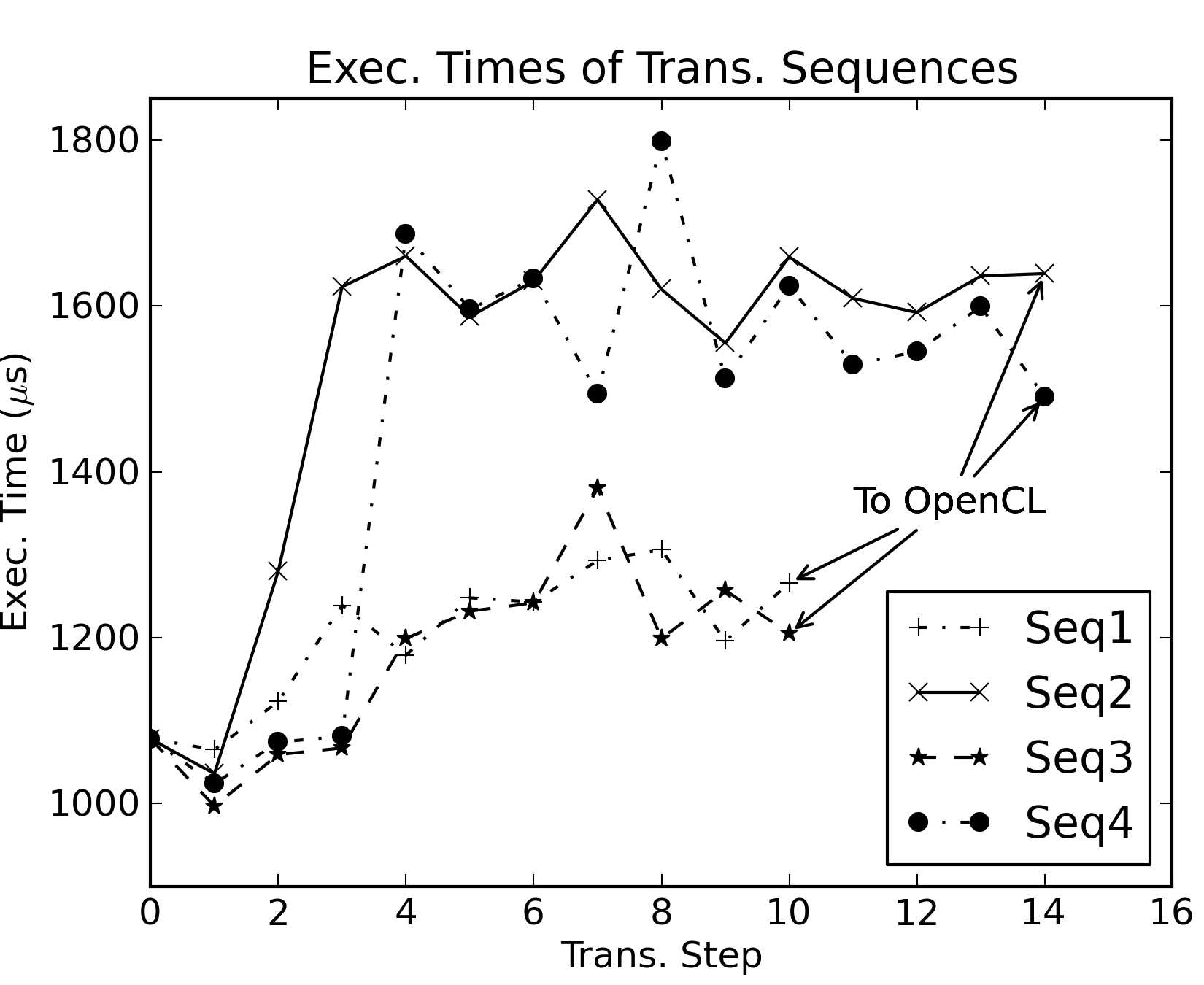}
\caption{Execution times for transformation sequences (on a CPU).}
\label{fig:seqExecTime}
\end{minipage}
\hfill
\begin{minipage}{0.45\textwidth}
\centering
\center
\includegraphics[width=0.9\textwidth]{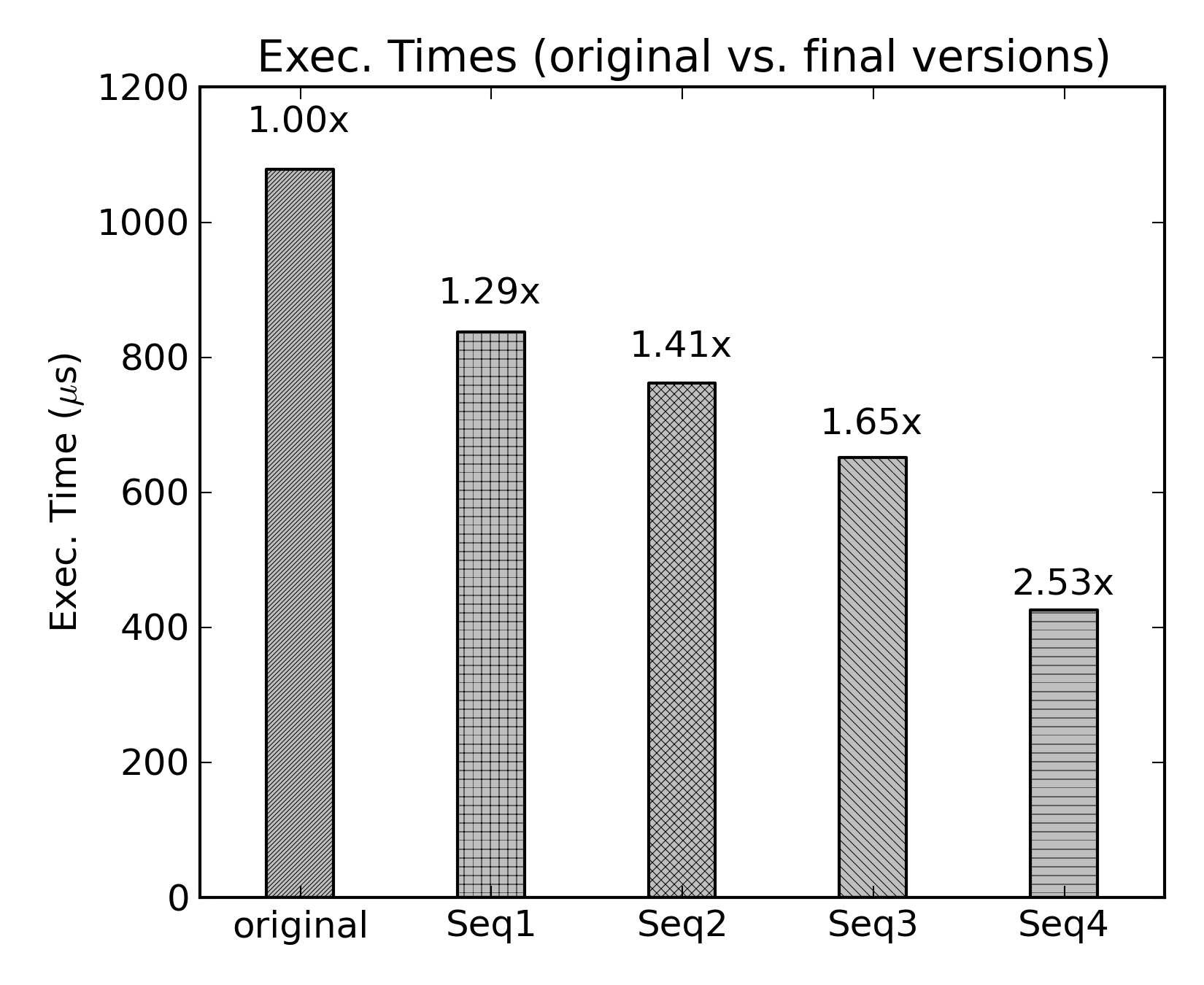}
\caption{Execution times for OpenCL versions (on a GPU). }
\label{fig:finalExecTime}
\end{minipage}
\end{figure}

Non-monotonicity has two sides: on one side while transforming (sequential) code to a shape better suited to some final platform, the efficiency observed in a CPU for a good sequence can increase and decrease.   On the other side, the final code which extracts the best performance on a CPU is not necessarily  the most efficient on the final platform.
In order to show this behavior we have identified four different
transformation sequences for a use case which performs image
compression using the discrete cosine transform.
%
Each of the identified sequences finally produces C code with a shape
adequate to be mechanically translated to OpenCL and executed on a
GPU.   We have measured the average execution time of 30 runs for each
intermediate state of each sequence. It can be seen
(Fig.~\ref{fig:seqExecTime}) that these execution times do not change monotonically. As expected, we also obtain different sequential codes \textit{ready} to be translated to OpenCL and with different execution times (still on a CPU).

As a second step, we have measured the performance of the OpenCL version generated from the final state of each sequence. Fig.~\ref{fig:finalExecTime} shows the execution times of the original sequential program and OpenCL versions as well as the speedup of each OpenCL version with respect to the sequential program (on top of each bar).  The OpenCL code obtained from sequence 4 is the fastest one, with a 2.53$\times$ speedup factor w.r.t. the initial code on a CPU.   However, looking at Fig.~\ref{fig:seqExecTime}, sequence 4 was not the fastest one on the CPU ---it was in fact the second slowest one.  Looking together at Fig.~\ref{fig:seqExecTime} and~\ref{fig:finalExecTime}, it is easy to see that there is no clear correlation between the non-functional property measured (execution time) of each final sequential code and the performance of the corresponding OpenCL version. 
Based on these results it can be concluded that an effective method must be used to discover and learn the uncorrelated relation between final sequential codes and parallel versions for a range of target platforms. We have decided to base our approach on \RL, since it is driven by final performance measurements rather than on intermediate values which can lead to sub-optimal results.

We have also performed a preliminary evaluation of \RL as a technique to learn and then guide a rule-based program transformation tool. 
We have selected four use cases and identified different
transformation sequences leading to code that can be mechanically
translated to OpenCL and executed on a GPU. These four cases and their
transformation sequences have been used as training set.  The cases in
the training set are the image compression program (\textit{compress}), mentioned before in Fig.~\ref{fig:seqExecTime} and~\ref{fig:finalExecTime}, an image filter that splits the different color channels of a given RGB input image in different images (\textit{rgbFilter}), the detection of edges in an image using a Sobel filter (\textit{edgeDetect}) and another one performing image segmentation given a threshold value (\textit{threshold}).

Once the training set is defined, the \RL process requires  tuning the two parameters in (\ref{eq:RL}): the learning rate ($\alpha$) and the discount factor ($\gamma$). For this purpose, an empirical study was performed in which parameter values leading to transformation sequences providing the fastest OpenCL versions were selected.
%
As an outcome, a value of 0.5 was used for $\alpha$ and 0.6 for $\gamma$.  Also, reward values have been chosen in order to give a higher reinforcement to those sequences leading to  final codes with better performance. In our example, the reward values used for the best sequences of each use case application are substantially higher, with a ratio of 1:100, with respect to the rest of transformation sequences.

After training,
three different use case applications were used as prediction set. We
have selected applications that share code patterns with the examples
in the prediction set.  This matches the idea that sets of
transformation rules can very naturally depend on the application
domain, and therefore it makes sense that the training sets also
belong to the same domain.
The applications in the prediction set were mechanically transformed according to the previously learned sequences and finally translated by hand into OpenCL.  Independently, OpenCL versions of the initial C code were manually written to provide a comparison with what a programmer would do.  The evaluation of the \RL approach was made
by comparing the performance of the hand-coded versions with that of
the versions mechanically generated. 


\begin{figure} 
  \begin{center}
\vspace{-1.0em}  
\includegraphics[width=0.45\textwidth]{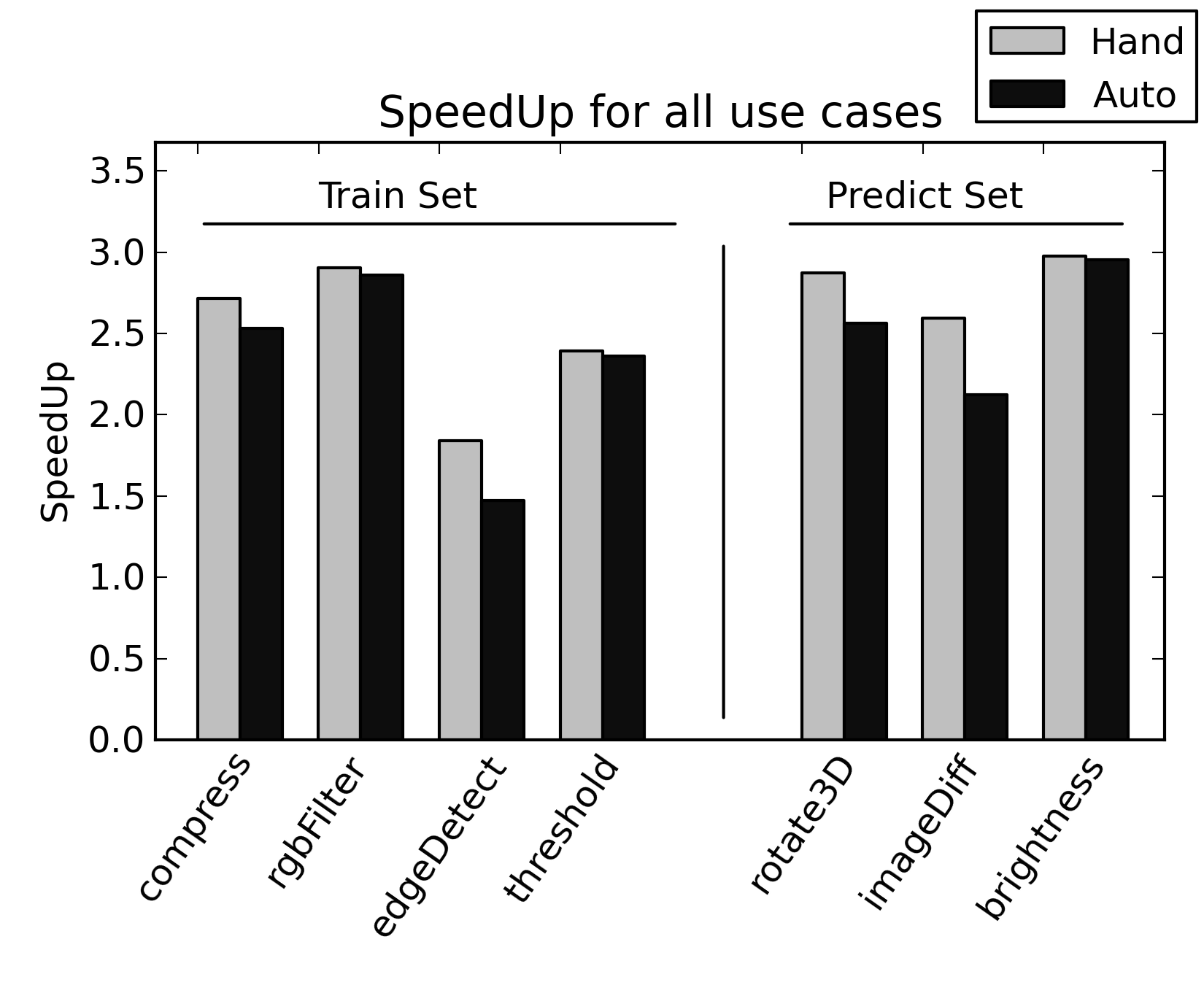}
  \end{center}
\vspace{-2.5em}
\caption{Speed up for train and prediction sets.}
\label{fig:allSpeedUp}
\end{figure}

Fig.~\ref{fig:allSpeedUp} shows the results obtained for  the training and prediction sets in terms of speedup of the OpenCL versions w.r.t.\ the original sequential program.  By comparing the results for the training and the prediction sets we can assess the behavior of the transformation system after the learning phase. Looking at results in Fig.~\ref{fig:allSpeedUp} we can see that the transformation sequences leads to code that provide acceleration factors comparable to those of the manually coded versions. Although this preliminary evaluation is based on a small sample, it shows that our approach seems promising in tackling the problem of deciding strategies for rule-based program transformation systems.

The results in this section show a preliminary, but very positive, evaluation for the generation of code for GPU platforms. We believe that the same approach can be followed in order to target different platforms which are present in heterogeneous systems. In this way, a separate \textit{state-action} table can be used for learning adequate transformation sequences for each target platform and rewards would be assigned based on the performance of the final code generated for each platform.

%
\savespacebeforesection
\section{Conclusions and Future Work}
\label{sec:conclusions}
\savespaceaftersection

We have proposed a \ML-based approach to learn heuristics for guiding the code transformation process of a rule-based program
transformation
system~\cite{tamarit15:padl-haskell_transformation,tamarit16:transformation-prole}. This
type of systems pose a number of problems, such as state explosion,
arising from the application of transformation rules in an arbitrary
order or the definition of a stop criteria.  For the latter we have
proposed the use of classification trees and for the former we have proposed a novel approach based on \RL.  We have also performed a preliminary evaluation of the approach, which provided promising results that demonstrate the suitability of the approach for this type of transformation systems.


As future work, we plan to continue expanding the set of use cases for
training the different \ML techniques used. As a consequence, we
expect to enrich the code features identified to obtain program
abstractions in order to capture new code patterns. 

New transformation sequences for the learning set can be automatically
synthesized by swapping commutative rules in existing sequences, doing
a sort of inverse of the partial order reduction used in model
checking algorithms.  While this may enrich the training set, it may
also introduce unwanted noise, and therefore we need to evaluate this
possibility carefully.

%
We also expect that the
increase in the training set will result in better prediction
outcomes, but it will also increase the complexity of efficiently
using the learning techniques. In the case of \RL, having more
learning cases results in a bigger state space.
 For that reason we foresee using common methods to reduce the number of states like clustering or principal component analysis techniques. We also plan to explore other features of \RL like the use of different learning rates for different states or transformation sequences in order to learn and converge faster towards transformed codes providing final versions with better performance. 

We also plan to define \textit{multi-objective} \RL rewards based on different non-functional properties like energy consumption, resource usage, etc. and even a combination of them. This future line would make it possible to define
transformation strategies that try to optimize different aspects and generate, for example, a final code among the fastest ones that consumes the least amount of energy. Also, the introduction of weights into the \textit{multi-objective} rewards would offer programmers the flexibility to select which non-functional property or set of properties they want to focus on for generating the final code.


\bibliographystyle{eptcs}
\bibliography{../../BiBTeX/hpc_transformations,../../BiBTeX/polca_refs,../../BiBTeX/machine_learning}

%
%

\end{document}